\documentclass[twocolumn,floatfix,showpacs,preprintnumbers,amsmath,amssymb,prb]{revtex4}

\usepackage{graphicx}
\usepackage{dcolumn}
\usepackage{bm}

\begin{document}

\input{epsf}

\title{Phase diagrams of the generalized spin-$\frac{1}{2}$ ladder under 
staggered field and dimerization: A renormalization group study}

\author{Y.-J. Wang}

\affiliation{Max-Planck-Institut f\"{u}r Physik Komplexer Systeme,
N\"{o}thnitzer Stra{\ss}e 38, 01187 Dresden, Germany} 

\date{\today}

\begin{abstract}

In the weak-coupling regime of the continuous theories, two sets of one-loop 
renormalization group equations are derived and solved to disclose the phase 
diagrams of the antiferromagnetic generalized two-leg spin-$\frac{1}{2}$ 
ladder under the effect of (I) a staggered external magnetic field and (II) an 
explicit dimerization. In model (I), the splitting of the SU(2)$_2$ 
critical line into U(1) and Z$_2$ critical surfaces is observed; while in 
model (II), two critical surfaces arising from their underlying critical lines 
with SU(2)$_2$ and Z$_2$ characteristics merge into an SU(2)$_1$ critical 
surface on the line where the model attains its highest symmetry. 

\end{abstract}

\pacs{75.10.Jm, 71.10.Pm, 64.60.Ak, 05.70.Jk}

\maketitle

\section{Introduction}               

The physics of quantum phase transitions in one-dimensional systems has 
attracted much attention in recent years because not only of its 
speciality---the conformal symmetries and universalities in (1+1)-critical 
models,\cite{gins88, aff88, gnt99} but also of its generality---the quantum 
critical phenomena in higher 
dimensions.\cite{cont01} A general question is posed as follows: If a
critical system defined in the ultraviolet limit is deformed by more
than one relevant perturbations, what is the fate of the system? Can
the criticality be reached again in the infrared limit because of the
``cancellation'' effect among these perturbations, thus a quantum phase
transition takes place? A typical model of this sort is the so-called
double-frequency sine-Gordon model,\cite{dm98,fgn00} where an Ising
transition was thought to be driven by two competing relevant
perturbations. The description of the phase transition requires a 
non-perturbative scheme being able to identify correctly those degrees
of freedom that remain massive and the low-energy ones that eventually
become critical and undergo the transition. 

The same strategy was extended to a more complicated non-Abelian 
case---a two-leg spin-$\frac{1}{2}$ Heisenberg antiferromagnetic ladder 
subject to a site-parity-breaking dimerization field.\cite{wn00} The non-trivial 
aspect of the non-perturbative approach adopted in this study of the 
strong-coupling limit of the model is that one has to preserve the SU(2) 
symmetry, which cannot be spontaneously broken, in the Ising model language 
which apparently has only a discrete symmetry. This hidden symmetry was realized 
when using very non-local duality transformations for the coupled Ising 
models. An effective low-energy Hamiltonian depicting the ensuing quantum 
phase transition was derived and the transmutation of all physical fields 
at the infrared fixed point was identified. It was unambiguously shown that 
a quantum phase transition occurs from the universality class of SU(2) level 2  
Wess-Zumino-Novikov-Witten (WZNW) model in the ultraviolet limit to the 
universality class of SU(2) level 1 WZNW model in the infrared limit. 

Another possible scenario was set up by introducing a staggered magnetic 
field, which explicitly breaks both SU(2) and bond-parity symmetries, to the 
generalized spin ladder.\cite{wefn02} Depending on which phase the model 
belongs to in absence of the staggered field, either U(1) or Z$_2$ criticality 
was predicted under the effect of the staggered field. An interesting 
new phase that interpolates between the Haldane spin liquid phase 
and the spontaneously dimerized phase was found beyond the transitions. 
The main features of this intermediate phase are: partially (transverse) 
coherent spin excitations and partially (longitudinal) non-vanishing string 
order parameter. 

As we can see, the physics of the generalized spin-$\frac{1}{2}$ ladder under 
the effect of the dimerization or the staggered field is very rich. In this 
paper, we shall scrutinize these two models in a more general perspective and 
try to understand the phase diagrams as a whole. Our main attention is to be 
paid to the overall topology of the phase diagrams; while the physical 
properties in various phases are only to be mentioned briefly if available. 

The paper is organized as follows. In Sec.~\ref{sec_mod}, we present a short 
overview of the continuous version of our models. In Sec.~\ref{sec_rgeq}, we 
derive two sets of renormalization group (RG) equations and central charge 
formulas. These equations are (numerically) solved and analyzed in 
Sec.~\ref{sec_sol} to bring about the phase diagrams of the continuous models. 
The implication to the ladder systems is summarized and concluded in 
Sec.~\ref{sec_sum}.

\section{Models and Their Continuous Field-Theoretic Mappings} \label{sec_mod} 

We consider a standard two-leg spin-$\frac{1}{2}$ antiferromagnetic ($J>0$) 
Heisenberg ladder ($J_\perp$) generalized to include a four-spin interaction 
($V$):\cite{nt97,km98} 
\begin{eqnarray}
H_{\text{gen}} &=& J \sum_{a=1,2} \sum_n \mbox{\boldmath $S$}_{a,n} \!\cdot\! 
\mbox{\boldmath $S$}_{a,n+1} 
+ J_\perp \sum_n \mbox{\boldmath $S$}_{1,n} \!\cdot\! 
\mbox{\boldmath $S$}_{2,n} \nonumber \\ 
&& + V \sum_n (\mbox{\boldmath $S$}_{1,n} \!\cdot\! \mbox{\boldmath $S$}_{1,n+1}) 
(\mbox{\boldmath $S$}_{2,n} \!\cdot\! \mbox{\boldmath $S$}_{2,n+1}) \;, 
\label{ham_gen} 
\end{eqnarray}
superimposed by either a staggered magnetic field, model (I), 
\begin{equation} 
H_{h} = h \sum_{a=1,2} \sum_n (-1)^n S_{a,n}^z \;, 
\label{ham_h}
\end{equation} 
or a $\pi$-phase (relative) dimerization, model (II), 
\begin{equation} 
H_{\Delta} = \Delta \sum_{a=1,2} \sum_n (-1)^{n+a} 
\mbox{\boldmath $S$}_{a,n} \!\cdot\! \mbox{\boldmath $S$}_{a,n+1} \;. 
\label{ham_dlt}
\end{equation} 
Throughout the paper, we treat the models only in the weak-coupling regime, 
providing 
\begin{equation} 
|J_\perp|, |V|, |h| \; (\text{or } |\Delta|) \ll J \;. \label{wccnd} 
\end{equation}
In this limit, the field-theoretic approach in continuum is of advance and 
power in describing the low-energy physics of the models. In fact, the 
resulting Hamiltonians are represented by four interacting Majorana (real) 
fermions or, equivalently, four coupled Ising models. 

According to Refs.~\onlinecite{snt96, nt97, wn00, wefn02}, we have the 
following mappings in terms of Majoranas ($\xi_{R,L}$) and Ising variables 
($\sigma, \mu$): 
\begin{equation}
{\cal H}_{\text{gen}} = {\cal H}_{\text{crit}} + {\cal H}_{\text{mass}} 
+ {\cal H}_{\text{marg}} \;, 
\end{equation}
where the critical theory 
\begin{equation} 
{\cal H}_{\text{crit}} = -\frac{iv}{2} \sum_{\mu=0}^3 \big(
\xi^\mu_R \partial_x \xi^\mu_R - \xi^\mu_L \partial_x \xi^\mu_L \big) 
\end{equation} 
having an O(4) symmetry is inherited from two decoupled Heisenberg chains 
with SU(2)$\times$SU(2) symmetry, $v\sim Ja$.\cite{note1} The mass terms result from part 
of the interchain coupling ($J_\perp$) and the four-spin interaction ($V$), 
both of which have the same scaling dimension, $d=1$: 
\begin{equation} 
{\cal H}_{\text{mass}} = -im_s\, \xi^0_R \xi^0_L - im_t\, 
\mbox{\boldmath $\xi$}_R \!\cdot\! \mbox{\boldmath $\xi$}_L \;, 
\end{equation} 
where the singlet (associated with $\xi^0$) and triplet 
($\mbox{\boldmath $\xi$} = (\xi^1, \xi^2, \xi^3)$) masses are related to 
the parameters of the lattice spin model by 
\begin{equation} 
m_s = -3 c J_\perp -c' V \;, \qquad 
m_t = c J_\perp -c' V \;, \label{mrjv} 
\end{equation} 
respectively. Here $c$ and $c'$ are some {\em positive} 
constants. The marginal terms are from the current-current part of the 
interchain coupling with scaling dimension $d=2$: 
\begin{equation} 
{\cal H}_{\text{marg}} = \frac{1}{2} g_1 ( \mbox{\boldmath $\xi$}_R \!\cdot\! 
\mbox{\boldmath $\xi$}_L)^2 + g_2 ( \mbox{\boldmath $\xi$}_R \!\cdot\! 
\mbox{\boldmath $\xi$}_L) (\xi_R^0 \xi_L^0) \;, 
\end{equation} 
where $g_1 = \frac{1}{2} J_\perp a -\gamma$ and $g_2 = -\frac{1}{2} J_\perp a 
-\gamma$, with $\gamma (>0)$ the marginally irrelevant coupling in the chains. 
Obviously, the symmetry of the continuous model ${\cal H}_{\text{gen}}$ is 
O(3)$\times$Z$_2$, corresponding to SU(2)$\times$Z$_2$ for the lattice model 
(\ref{ham_gen}) (Z$_2$ arising from the interchange of two chains),  
unless $J_\perp =0$ when the highest symmetry of O(4) is reached, 
corresponding to SU(2)$\times$SU(2). 

The field-theoretic counterparts of the staggered field (\ref{ham_h}) and the 
dimerization (\ref{ham_dlt}) are represented, respectively, by 
\begin{eqnarray} 
{\cal H}_h &=& (h/\alpha) \sigma_1 \sigma_2 \mu_3 \mu_0 \;, \\ 
{\cal H}_\Delta &=& (\Delta/\alpha) \sigma_1 \sigma_2 \sigma_3 \sigma_0 \;,  
\end{eqnarray} 
where $\alpha$ with a magnitude order of the lattice constant is the 
short-distance cut-off of the theory. While ${\cal H}_\Delta$ keeps the O(3) 
invariance for the triplet sector intact, ${\cal H}_h$ breaks this symmetry down to 
O(2)$\times$Z$_2$. This is a reflection of the oriented field (\ref{ham_h}) 
lowers SU(2) to U(1)$\times$Z$_2$. 

In what follows, we shall address model (I): ${\cal H}_{\text{gen}} + 
{\cal H}_h$, and model (II): ${\cal H}_{\text{gen}} + {\cal H}_\Delta$, in RG 
approach. Our purpose aims at drawing the phase diagrams by analyzing the 
infrared properties of the RG equations.

\section{Derivation of the RG Equations} \label{sec_rgeq}

It is rather straightforward to establish the one-loop RG equations out of 
the operator product expansions\cite{gins88} (OPE's) between the perturbative 
operators.\cite{cardy88} For a critical model perturbed by 
\begin{equation} 
{\cal H}_{\text{pert}} = \frac{v}{\pi\alpha^2} \sum_i \lambda_i {\cal O}_i \;, 
\end{equation} 
where $\lambda_i$ are dimensionless coupling constants of the corresponding 
dimensionless operators ${\cal O}_i$ (with scaling dimension $d_i$), if 
${\cal O}_i$ are normalized in such a way that 
\begin{equation} 
\langle {\cal O}_i(z,\bar{z}) {\cal O}_j(0,0) \rangle 
= \delta_{ij} \bigg( \frac{\alpha}{|z|} \bigg)^{2d_i} \;, 
\label{norm}  
\end{equation} 
($z= v\tau + ix$ being the time-space in complex coordinates) and the 
short-distance OPE's satisfy 
\begin{equation} 
{\cal O}_i(z,\bar{z}) {\cal O}_j(0,0) 
\sim {c_{ij}}^k \bigg( \frac{\alpha}{|z|} \bigg)^{d_i+d_j-d_k} 
{\cal O}_k(0,0) \;,  
\end{equation}
the one-loop RG equations for $\lambda_i$ are readily given by 
\begin{equation} 
\dot{\lambda}_k \equiv \frac{d \lambda_k}{d \ln \!L} 
= (2-d_k) \lambda_k - \sum_{ij} 
{c_{ij}}^k \lambda_i \lambda_j + O(\lambda^3) \;. \label{rgeq}
\end{equation} 

In addition, the central charge satisfying Zamo\-lod\-chi\-kov's decreasing 
theorem\cite{zam86} is perturbatively evaluated to this order to be 
\begin{eqnarray} 
C(\{\lambda\}) &=& C_{\text{UV}} - 3 \sum_k (2-d_k) \lambda_k^2 \nonumber \\ 
&& + 2 \sum_{ijk} {c_{ij}}^k \lambda_i \lambda_j \lambda_k 
+ O(\lambda^4) \;, \label{cnch} 
\end{eqnarray} 
where $C_{\text{UV}}$ is the central charge of the unperturbed critical system 
(the ultraviolet fixed point). In the present case, $C_{\text{UV}} =2$ is the 
central charge of two decoupled chains. 

Under normalization (\ref{norm}), the OPE coefficients ${c_{ij}}^k$ are 
related to the three-point function $\langle {\cal O}_i {\cal O}_j {\cal O}_k 
\rangle$, and thus are complete symmetric among the indices 
($ijk$).\cite{gins88} Not only this procedure simplifies the calculations, 
but also is it essential to give rise to a correct form for the central charge 
(\ref{cnch}).

\subsection{RG equations for model (I)} 

Since for this model, 
\begin{equation} 
{\cal H}_{\text{pert}}^{\text{(I)}} = {\cal H}_{\text{mass}} 
+ {\cal H}_{\text{marg}} + {\cal H}_{h} \;, 
\end{equation} 
and noticing ${\cal H}_{h}$ breaks the O(3) symmetry in the triplet sector, 
we have to divide the triplet mass $m_t$ and the marginal coupling constants 
$g_1$ and $g_2$ into doublets and singlets. Denoting the mass bilinear (or 
energy density) of the Ising model, $\varepsilon_\mu = i\xi_R^\mu \xi_L^\mu$, 
we write down: 
\begin{eqnarray} 
{\cal H}_{\text{pert}}^{\text{(I)}} &=& 
-m_t^d (\varepsilon_1 + \varepsilon_2) 
-m_t^s \varepsilon_3 - m_s \varepsilon_0 \nonumber \\ 
&& - g_1^d \varepsilon_1 \varepsilon_2
- g_1^s (\varepsilon_1 + \varepsilon_2) \varepsilon_3 \nonumber \\ 
&& - g_2^d (\varepsilon_1 + \varepsilon_2) \varepsilon_0 
- g_2^s \varepsilon_3 \varepsilon_0 \nonumber \\ 
&& + (h/\alpha) \sigma_1 \sigma_2 \mu_3 \mu_0 \nonumber \\ 
&=& \frac{v}{\pi\alpha^2} \sum_{i=1}^8 \lambda_i {\cal O}_i \;, 
\end{eqnarray} 
where the dimensionless operators in both fermion and Abelian boson 
representations (``$+/-$'' associated with sectors $(1,2)/(3,0)$, 
respectively): 
\begin{eqnarray} 
&& {\cal O}_1 = -\sqrt{2} \pi\alpha (\varepsilon_1 + \varepsilon_2) 
= -\sqrt{2} \cos\sqrt{4\pi} \Phi_+ \;, \nonumber \\ 
&& {\cal O}_2 = - 2\pi\alpha \varepsilon_3 
= -\cos\sqrt{4\pi} \Phi_- + \cos\sqrt{4\pi} \Theta_- \;, \nonumber \\ 
&& {\cal O}_3 = - 2\pi\alpha \varepsilon_0 
= -\cos\sqrt{4\pi} \Phi_- - \cos\sqrt{4\pi} \Theta_- \;; \nonumber \\ 
&& {\cal O}_4 = -(2\pi\alpha)^2 \varepsilon_1 \varepsilon_2 
= 4\pi\alpha^2 \bar{\partial} \Phi_+ \partial \Phi_+ \;, \nonumber \\ 
&& {\cal O}_5 = -2\sqrt{2} (\pi\alpha)^2 (\varepsilon_1 + \varepsilon_2) 
\varepsilon_3 \nonumber \\ 
&& \hspace{1.5em} = -\sqrt{2} \cos\sqrt{4\pi} \Phi_+ 
\big(\cos\sqrt{4\pi} \Phi_- - \cos\sqrt{4\pi} \Theta_- \big) \;, \nonumber \\ 
&& {\cal O}_6 = -2\sqrt{2} (\pi\alpha)^2 (\varepsilon_1 + \varepsilon_2) 
\varepsilon_0 \nonumber \\ 
&& \hspace{1.5em} = -\sqrt{2} \cos\sqrt{4\pi} \Phi_+ 
\big(\cos\sqrt{4\pi} \Phi_- + \cos\sqrt{4\pi} \Theta_- \big) \;, \nonumber \\
&& {\cal O}_7 = -(2\pi\alpha)^2 \varepsilon_3 \varepsilon_0 
= 4\pi\alpha^2 \bar{\partial} \Phi_- \partial \Phi_- \;; \nonumber \\ 
&& {\cal O}_8 = 2 \sigma_1 \sigma_2 \mu_3 \mu_0 
= 2 \sin\sqrt{\pi} \Phi_+ \cos\sqrt{\pi} \Phi_- \;. 
\end{eqnarray} 
The scaling dimensions, $d_{1,2,3} = 1$, $d_{4,5,6,7} = 2$, and 
$d_8 = \frac{1}{2}$. Accordingly, the dimensionless coupling constants: 
\begin{eqnarray} 
&& \lambda_1 \equiv \frac{\eta_1}{\sqrt{2}} = \frac{m_t^d \alpha}{\sqrt{2}v} 
\;, \qquad 
\lambda_2 \equiv \frac{\eta_2}{2} = \frac{m_t^s \alpha}{2v} \;, \nonumber \\ 
&& \lambda_3 \equiv \frac{\eta_3}{2} = \frac{m_s \alpha}{2v} \;; \nonumber \\ 
&& \lambda_4 \equiv \frac{\eta_4}{4} = \frac{g_1^d}{4\pi v} \;, \qquad 
\lambda_5 \equiv \frac{\eta_5}{2\sqrt{2}} = \frac{g_1^s}{2\sqrt{2}\pi v} \;, 
\nonumber \\
&& \lambda_6 \equiv \frac{\eta_6}{2\sqrt{2}} = \frac{g_2^d}{2\sqrt{2}\pi v} 
\;, \qquad 
\lambda_7 \equiv \frac{\eta_7}{4} = \frac{g_2^s}{4\pi v} \;; \nonumber \\ 
&& \lambda_8 \equiv \frac{\eta_8}{2} = \frac{\pi h}{2v} \;. \label{cpct} 
\end{eqnarray} 

It is easy to calculate all the OPE coefficients by using either 
fermion or boson representation. For example, we find\cite{note2}   
\begin{eqnarray}
&& \hspace{-2em} {\cal O}_8(z,\bar{z}) {\cal O}_8 \sim \frac{\alpha}{|z|} 
+ \frac{1}{\sqrt{2}} {\cal O}_1 -\frac{1}{2} ( {\cal O}_2 + {\cal O}_3 ) 
\nonumber \\ 
&& + \bigg( \frac{\alpha}{|z|} \bigg)^{-1} 
\bigg[ \frac{1}{2\sqrt{2}} ( {\cal O}_5 + {\cal O}_6 ) 
- \frac{1}{4} ( {\cal O}_4 + {\cal O}_7 ) \bigg] \;. \label{o8ope} 
\end{eqnarray} 
Therefore, we read off 
\begin{eqnarray}
&& {c_{88}}^1 = \frac{1}{\sqrt{2}} \;, \quad 
{c_{88}}^2 = {c_{88}}^3 = -\frac{1}{2} \;, \nonumber \\ 
&& {c_{88}}^5 = {c_{88}}^6 = \frac{1}{2\sqrt{2}} \;, \quad 
{c_{88}}^4 = {c_{88}}^7 = -\frac{1}{4} \;. 
\end{eqnarray} 
The algebra in this way is closed, and we list all the coefficients in 
Table~\ref{tab_coe}. 
\begin{table}[ht] 
\caption{OPE coefficients ${c_{ij}}^k$ (given in the parentheses, in front are 
indices $k$).}
\begin{tabular}{c|cccccccc}
\hline\hline
$i\backslash j$ & 1 & 2 & 3 & 4 & 5 & 6 & 7 & 8 \\ \colrule
1 & 4(-1) & 5(-1) & 6(-1) & 1(-1) & 2(-1) & 3(-1) 
& & 8($\frac{1}{\sqrt{2}}$) \\ 
2 & 5(-1) & & 7(-1) & & 1(-1) & & 3(-1) & 8(-$\frac{1}{2}$) \\ 
3 & 6(-1) & 7(-1) & & & & 1(-1) & 2(-1) & 8(-$\frac{1}{2}$) \\ 
4 & 1(-1) & & & & 5(-1) & 6(-1) & & 8(-$\frac{1}{4}$) \\ 
5 & 2(-1) & 1(-1) & & 5(-1) & 4(-1) & 7(-1) 
& 6(-1) & 8($\frac{1}{2\sqrt{2}}$) \\
6 & 3(-1) & & 1(-1) & 6(-1) & 7(-1) & 4(-1) & 
5(-1) & 8($\frac{1}{2\sqrt{2}}$) \\  
7 & & 3(-1) & 2(-1) & & 6(-1) & 5(-1) & 
& 8(-$\frac{1}{4}$) \\ 
8 & 8($\frac{1}{\sqrt{2}}$) & 8(-$\frac{1}{2}$) & 8(-$\frac{1}{2}$) 
& 8(-$\frac{1}{4}$) & 8($\frac{1}{2\sqrt{2}}$) & 8($\frac{1}{2\sqrt{2}}$) 
& 8(-$\frac{1}{4}$) & (${c_{88}}^k$)\footnote{ $\;{c_{88}}^k 
= 1(\frac{1}{\sqrt{2}}), \, 2(-\frac{1}{2}), \, 3(-\frac{1}{2}), 
\, 4(-\frac{1}{4}), \, 5(\frac{1}{2\sqrt{2}}), \, 6(\frac{1}{2\sqrt{2}}), 
\, 7(-\frac{1}{4})$.} \\ 
\hline\hline
\end{tabular} 
\label{tab_coe} 
\end{table} 

Now, it follows from Eq.~(\ref{rgeq}) that the RG equations for the 
couplings $\eta$ (instead of $\lambda$): 
\begin{eqnarray}
\dot{\eta}_1 &=& \eta_1 + \frac{1}{2} (\eta_1 \eta_4 + \eta_2 \eta_5 
+ \eta_3 \eta_6 ) - \frac{1}{4} \eta_8^2 \;, \nonumber \\ 
\dot{\eta}_2 &=& \eta_2 + \frac{1}{2} ( 2 \eta_1 \eta_5 
+ \eta_3 \eta_7 ) + \frac{1}{4} \eta_8^2 \;, \nonumber \\ 
\dot{\eta}_3 &=& \eta_3 + \frac{1}{2} ( 2 \eta_1 \eta_6
+ \eta_2 \eta_7 ) + \frac{1}{4} \eta_8^2 \;; \nonumber \\ 
\dot{\eta}_4 &=& 2 \eta_1^2 + \frac{1}{2} (\eta_5^2 + \eta_6^2)   
+ \frac{1}{4} \eta_8^2 \;, \nonumber \\ 
\dot{\eta}_5 &=& 2 \eta_1 \eta_2 + \frac{1}{2} (\eta_4 \eta_5 + 
\eta_6 \eta_7 ) -  \frac{1}{4} \eta_8^2 \;, \nonumber \\ 
\dot{\eta}_6 &=& 2 \eta_1 \eta_3 + \frac{1}{2} (\eta_4 \eta_6 +   
\eta_5 \eta_7 ) -  \frac{1}{4} \eta_8^2 \;, \nonumber\\ 
\dot{\eta}_7 &=& 2 \eta_2 \eta_3 + \eta_5 \eta_6  
+ \frac{1}{4} \eta_8^2 \;; \nonumber \\ 
\dot{\eta}_8 &=& \frac{3}{2} \eta_8 - \eta_1 \eta_8 
+ \frac{1}{2} (\eta_2 + \eta_3) \eta_8  \nonumber \\
&&  - \frac{1}{4} (\eta_5 + \eta_6) \eta_8  
+ \frac{1}{8} (\eta_4 + \eta_7) \eta_8 \;. \label{rgm1} 
\end{eqnarray} 
The perturbative central charge (\ref{cnch}) for the present model in 
terms of $\eta$ is 
\begin{eqnarray}
C(\{\eta\}) &=& 2 - \frac{3}{4} \Big( 2\eta_1^2 + \eta_2^2 + \eta_3^2 
+ \frac{3}{2} \eta_8^2 \Big) \nonumber \\ 
&& -\frac{3}{2} \Big(\eta_1 \eta_2 \eta_5 + \eta_1 \eta_3 \eta_6 
+ \frac{1}{2} \eta_2 \eta_3 \eta_7 + \frac{1}{4} \eta_5 \eta_6 \eta_7 \Big) 
\nonumber \\  
&& -\frac{3}{16} (4\eta_1^2 + \eta_5^2 + \eta_6^2) \eta_4 \nonumber \\  
&& + \frac{3}{8} \Big( 2 \eta_1 - \eta_2 - \eta_3 - \frac{1}{4} \eta_4 
\nonumber \\  
&& \qquad + \frac{1}{2} \eta_5 + \frac{1}{2} \eta_6 
- \frac{1}{4} \eta_7 \Big) \eta_8^2 \;. 
\end{eqnarray}

\subsection{RG equations for model (II)} 

Comparing with the previous model, the only difference now is replacing 
${\cal H}_h$ by ${\cal H}_\Delta$, i.e., 
\begin{eqnarray}
&& {\cal O}_8 = 2 \sigma_1 \sigma_2 \sigma_3 \sigma_0 
= 2 \sin\sqrt{\pi} \Phi_+ \sin\sqrt{\pi} \Phi_- \;, \\ 
&& \lambda_8 \equiv \frac{\eta_8}{2} = \frac{\pi\Delta}{2v} \;,  
\end{eqnarray} 
and Eq.~(\ref{o8ope}) is modified to be 
\begin{eqnarray}
&& \hspace{-2em} {\cal O}_8(z,\bar{z}) {\cal O}_8 \sim \frac{\alpha}{|z|} 
+ \frac{1}{\sqrt{2}} {\cal O}_1 +\frac{1}{2} ( {\cal O}_2 + {\cal O}_3 ) 
\nonumber \\ 
&& - \bigg( \frac{\alpha}{|z|} \bigg)^{-1} 
\bigg[ \frac{1}{2\sqrt{2}} ( {\cal O}_5 + {\cal O}_6 ) 
+ \frac{1}{4} ( {\cal O}_4 + {\cal O}_7 ) \bigg] \;,   
\end{eqnarray} 
yielding 
\begin{equation} 
\begin{array}{rl} {c_{88}}^k = & 1\big(\frac{1}{\sqrt{2}}\big)\,, \; 
2\big(\frac{1}{2}\big)\,, \; 3\big(\frac{1}{2}\big)\,, \\ 
& 4\big(\!-\!\frac{1}{4}\big)\,, \; 5\big(\!-\!\frac{1}{2\sqrt{2}}\big)\,, \; 
6\big(\!-\!\frac{1}{2\sqrt{2}}\big)\,, \; 7\big(\!-\!\frac{1}{4}\big) 
\;. \end{array} 
\end{equation} 

However, as we have known, the symmetry of model (II) is higher than that of 
model (I)---the O(3) symmetry in the triplet sector [i.e., the SU(2) symmetry 
in the original lattice model] should remain. In other 
words, the division of the triplet into doublet and singlet is futile in the 
present case. This fact is reflected in the RG equations also. From 
Eqs.~(\ref{rgm1}), by reversing the signs of the coefficients involving 
${c_{88}}^2$, ${c_{88}}^3$, ${c_{88}}^5$, and ${c_{88}}^6$, three of the eight 
equations are redundant, implying $\eta_1=\eta_2$, $\eta_4=\eta_5$, and 
$\eta_6=\eta_7$. As a result, the RG equations degenerate to 
\begin{eqnarray}
\dot{\eta}_1 &=& \eta_1 + \frac{1}{2} ( 2 \eta_1 \eta_4 
+ \eta_3 \eta_6 ) - \frac{1}{4} \eta_8^2 \;, \nonumber \\ 
\dot{\eta}_3 &=& \eta_3 + \frac{3}{2} \eta_1 \eta_6
- \frac{1}{4} \eta_8^2 \;; \nonumber \\ 
\dot{\eta}_4 &=& 2 \eta_1^2 + \frac{1}{2} (\eta_4^2 + \eta_6^2)   
+ \frac{1}{4} \eta_8^2 \;, \nonumber \\ 
\dot{\eta}_6 &=& 2 \eta_1 \eta_3 + \eta_4 \eta_6   
+ \frac{1}{4} \eta_8^2 \;; \nonumber\\ 
\dot{\eta}_8 &=& \frac{3}{2} \eta_8 - \frac{1}{2} (3 \eta_1 + \eta_3) \eta_8 
+ \frac{3}{8} (\eta_4 + \eta_6) \eta_8 \;, \label{rgm2} 
\end{eqnarray} 
and the central charge to 
\begin{eqnarray}
C(\{\eta\}) &=& 2 - \frac{3}{4} \Big( 3\eta_1^2 + \eta_3^2 
+ \frac{3}{2} \eta_8^2 \Big) \nonumber \\ 
&& -\frac{3}{2} \Big(\eta_1^2 \eta_4 + \frac{3}{2} \eta_1 \eta_3 \eta_6 
+ \frac{1}{4} \eta_4 \eta_6^2 \Big) 
\nonumber \\  
&& -\frac{3}{16} (4\eta_1^2 + \eta_4^2 + \eta_6^2) \eta_4 \nonumber \\  
&& + \frac{3}{8} \Big( 3 \eta_1 + \eta_3 
- \frac{3}{4} \eta_4 - \frac{3}{4} \eta_6 \Big) \eta_8^2 \;.  
\end{eqnarray} 

\section{Solutions to the RG Equations and the Emerging Phase Diagrams} 
\label{sec_sol}

To analyze the infrared behaviors of our present models, in this section, 
we numerically integrate these two sets of RG Eqs.~(\ref{rgm1}) and (\ref{rgm2}) 
to depict generic phase diagrams for the spin ladder under a staggered field 
or dimerization. On account of the fact that for a system in a massive regime, 
the role of marginal coupling is usually exhausted  by renormalizing the mass 
and velocity, it is reasonable to take no thought of these marginal 
perturbations at the outset and argue that the overall qualitative feature 
remains correct although the position and shape of phase boundary may not be 
exact. Let us begin with a known system---the generalized spin 
ladder (\ref{ham_gen})\cite{nt97}---as an example. A variant of this model 
includes four-spin ring exchange, which is believed to be relevant to some 
exotic properties in real ladder materials. \cite{breh99,hn02,mvm02}

In the following we denote the bare ultraviolet couplings by 
$\eta^{(0)} \equiv \eta(L=1)$ and the renormalized infrared couplings by 
$\eta^{(\infty)} \equiv \eta(L=\infty)$.  

\subsection{Phase diagram of the generalized spin ladder} 

\begin{figure}[ht]
\begin{picture}(0,200)
\leavevmode\centering\includegraphics{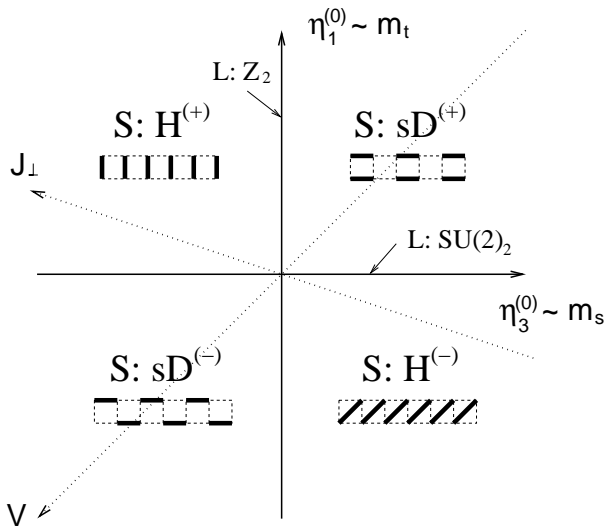}
\end{picture}
\caption{Phase diagram of the generalized spin ladder. There exist two 
critical lines L:~SU(2)$_2$ and L:~Z$_2$, and four massive sheets 
S:~H$^{(\pm)}$ and S:~sD$^{(\pm)}$. Notice the dimers in the Haldane phases 
are only symbolic rather than a true order.} 
\label{fig_pd_gl} 
\end{figure}

In this case, the RG equations are trivially $\dot{\eta}_1 = \eta_1$ and 
$\dot{\eta}_3 = \eta_3$. Unless the initial value (bare coupling) vanishes, 
$\eta_1^{(0)}=0$ or $\eta_3^{(0)}=0$, the system is always renormalized to 
some strong-coupling massive phase as $L\to \infty$. Because of relations 
in Eqs.~(\ref{cpct}), $\eta_1^{(0)} = \frac{m_t \alpha}{v}$ and $\eta_3^{(0)} 
= \frac{m_s \alpha}{v}$, we conclude that $m_t =0$ and $m_s =0$ are phase 
transition lines on which the model becomes critical. They belong to the 
universality classes of critical SU(2)$_2$ WZNW model and Z$_2$ Ising model, 
respectively.\cite{nt97} As a base for our further exploration, we draw the phase 
diagram of the generalized spin ladder (\ref{ham_gen}) in Fig.~\ref{fig_pd_gl}. 

According to Ref.~\onlinecite{nt97}, there are four massive phases divided by 
two critical lines. The second and fourth quadrants of the figure are two 
Haldane (H$^{(\pm)}$) spin liquid phases, and the first and third quadrants 
belong to spontaneously dimerized (sD$^{(\pm)}$) phases. The ``$+/-$'' sectors 
are actually related by duality of reversing the signs of the masses. 
The axis of abscissas represents a critical 
model with an SU(2)$_2$ WZNW universality class (L:SU(2)$_2$), and the axis of 
ordinates a Z$_2$ (Ising) one (L:Z$_2$). Although the Haldane phases and the 
dimer phases share the feature of non-vanishing topological string order 
parameter in common and thermodynamically they are indistinguishable, the 
symmetry of the ground state and the behavior of the dynamical susceptibility 
are completely different. 

(i) S:~H$^{(+)}$ phase [$\eta_1^{(\infty)}=+\infty$, $\eta_3^{(\infty)}=-\infty$]: 
Neither parity (site or bond) symmetry nor translational symmetry is broken. 
There is a coherent peak for the dynamical spin susceptibility at $q_\perp=\pi$, 
$q\sim \pi$, $\omega \sim |m_t|$, characterizing coherent $S=1$ single-magnon 
excitations. 

(ii) S:~H$^{(-)}$ phase [$\eta_1^{(\infty)}=-\infty$, $\eta_3^{(\infty)}=+\infty$]: 
No symmetry breaking (except for the hidden string order). The peak is at 
$q_\perp=0$, $q\sim \pi$, $\omega \sim |m_t|$ instead. 

(iii) S:~sD$^{(+)}$ phase [$\eta_1^{(\infty)}=+\infty$, $\eta_3^{(\infty)}=+\infty$]: 
The discrete site-parity and translational symmetries are spontaneously 
broken. The corresponding order parameter is $\langle (-1)^n 
(\mbox{\boldmath $S$}_{1,n} \!\cdot\! \mbox{\boldmath $S$}_{1,n+1} 
+ \mbox{\boldmath $S$}_{2,n} \!\cdot\! \mbox{\boldmath $S$}_{2,n+1}) \rangle 
\ne 0$ (0-phase). No coherent peak in the spin susceptibility (square-root 
singularity only). 

(iv) S:~sD$^{(-)}$ phase [$\eta_1^{(\infty)}=-\infty$, $\eta_3^{(\infty)}=-\infty$]: 
The discrete symmetries are broken with an order parameter $\langle (-1)^n 
(\mbox{\boldmath $S$}_{1,n} \!\cdot\! \mbox{\boldmath $S$}_{1,n+1} 
- \mbox{\boldmath $S$}_{2,n} \!\cdot\! \mbox{\boldmath $S$}_{2,n+1}) \rangle 
\ne 0$ ($\pi$-phase). No coherent peak either. 

We also show in the figure the directions of the lattice model parameters 
$J_\perp$ and $V$ via relation (\ref{mrjv}), although the exact scales for the 
coordinates are not known. We find that the standard ladder with positive, 
antiferro- (negative, ferro-) $J_\perp$ is in Haldane H$^{(+)}$ (H$^{(-)}$) 
phase. The pure four-spin coupling model with $J_\perp=0$ is in the sD$^{(-)}$ 
phase when $V>0$ or in the sD$^{(+)}$ phase when $V<0$. This is a special case 
with a higher symmetry. 

It is also interesting to check in RG sense the flow of the central charge. 
For this simple model we have 
\begin{equation}
C = 2 - \frac{3}{4} ( 3\eta_1^2 + \eta_3^2 ) \;. 
\end{equation} 
Considering this is a perturbative one-loop result and should not be taken 
seriously when $\eta$ approaches unity, we assume that we can mimic the true 
fixed point property by introducing a cut-off in $\eta$. If we are allowed 
to choose this cut-off to be $\sqrt{\frac{2}{3}}$, the correct central charges 
for both SU(2)$_2$ and Z$_2$ can be recovered, since for the former when 
$(\eta_1^*, \eta_3^*)=(0, \sqrt{\frac{2}{3}})$, 
$C^* = \frac{3}{2}$, and for the latter 
when $(\eta_1^*, \eta_3^*)=(\sqrt{\frac{2}{3}},0)$, 
$C^* = \frac{1}{2}$. The existence of the SU(2)$_2$ criticality was recently 
confirmed by a direct estimate of the numerical value of the central 
charge.\cite{hn02}

\subsection{Phase diagram of model (I)} 

Model (I) is the generalized spin ladder (\ref{ham_gen}) in a staggered field 
(\ref{ham_h}). The RG equations regarding the relevant perturbations now become 
\begin{eqnarray}
\dot{\eta}_1 &=& \eta_1 - \frac{1}{4} \eta_8^2 \;, \nonumber \\ 
\dot{\eta}_2 &=& \eta_2 + \frac{1}{4} \eta_8^2 \;, \nonumber \\ 
\dot{\eta}_3 &=& \eta_3 + \frac{1}{4} \eta_8^2 \;, \nonumber \\ 
\dot{\eta}_8 &=& \frac{3}{2} \eta_8 - \eta_1 \eta_8 
+ \frac{1}{2} (\eta_2 + \eta_3) \eta_8 \;. \label{dfeq_m1}
\end{eqnarray} 
Evidently, the equations are invariant under the transformation 
$\eta_8 \to -\eta_8$, which corresponds to $h\to -h$ in the original 
spin model. We need only deal with the positive case. At first glance 
the system has two mathematical fixed points: (a) $(\eta_1^*, \eta_2^*, 
\eta_3^*, \eta_8^*) = (0,0,0,0)$ and (b) $(\frac{3}{4}, -\frac{3}{4}, 
-\frac{3}{4}, \sqrt{3})$. The one at the origin is an unstable fixed 
point recounting two decoupled chains as our starting point; while the 
fixed point (b) is fallacious. This is due to the fact that $\eta_1$ 
and $\eta_2$ are not completely independent, i.e., initially $\eta_1^{(0)} 
= \eta_2^{(0)} = \frac{m_t\alpha}{v}$, rendering the fixed point (b) 
physically unaccessible. However, the values in (b) do have full play in the  
the structure of the phase diagram. 

\begin{figure}[ht]
\begin{picture}(0,130)
\leavevmode\centering\includegraphics{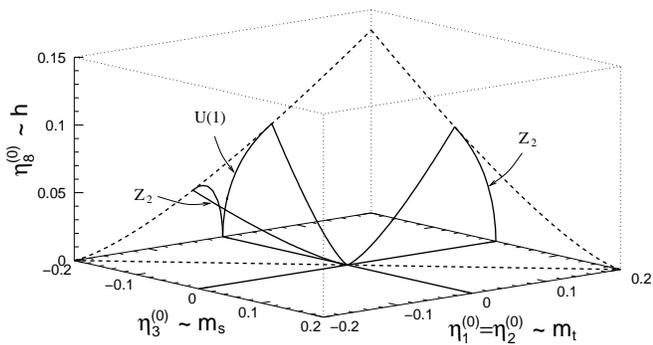}
\end{picture}
\caption{Phase diagram of the generalized spin ladder in a staggered field. 
The massive phases occupying different blocks are separated by interfaces 
either second-ordered (enclosed by solid line) or with the transition nature 
unknown from the RG equations alone (dashed line).} 
\label{fig_pd_md1} 
\end{figure}
\begin{figure}[ht]
\begin{picture}(0,260)
\leavevmode\centering\includegraphics{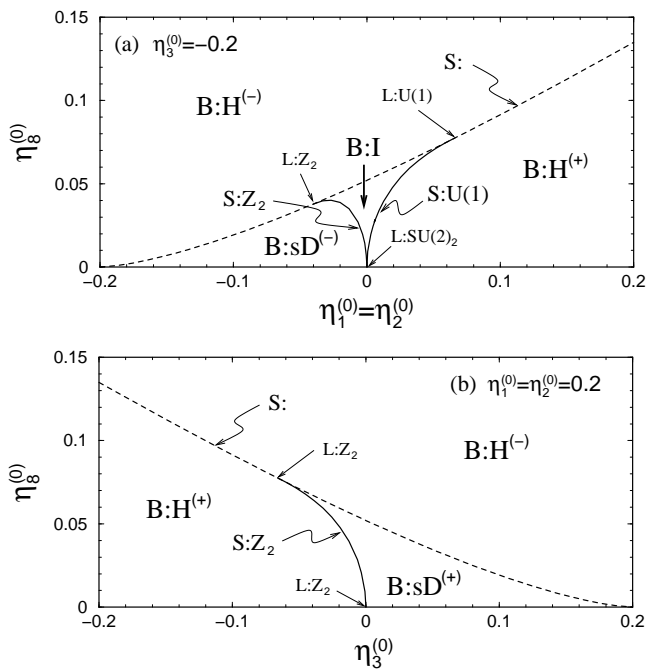}
\end{picture}
\caption{Sectional plot of the phase diagram of the generalized spin 
ladder in a staggered field: (a) $\eta_3^{(0)}$ is a constant as $-0.2$. 
(b) $\eta_1^{(0)} = \eta_2^{(0)}$ is a constant as $0.2$.} 
\label{fig_md1_sec} 
\end{figure}

 
The emerging phase diagram of model (I) is shown in Fig.~\ref{fig_pd_md1}. 
It is notable that all the interesting phases lie in the first three quadrants 
of the base and the fourth quadrant is in some sense rather barren. This is 
controlled by the ``fixed point'' (b) mentioned above. To make the structure 
clearer, we anatomically plot figures in two intersecting planes (see 
Fig.~\ref{fig_md1_sec}). Various phases are labeled in the figures, where 
``B'' denotes a block or bulk space that is always massive, ``S'' stands for 
a surface or sheet as phase boundary separating two different phases, which 
can be critical or first-ordered, and ``L'' is an intersectional line of two 
sheets. 

Now let us interpret in some detail the obtained phase diagram. We take the 
example with a parameter range where two consecutive phase transitions occur 
when increasing the staggered field---one belongs to the U(1) universality 
class; the other's property is not determined by the RG theory itself 
(probably first-ordered). Suppose a generalized spin ladder offers such a 
kind of parameters: 
\begin{equation} 
\begin{array}{l} 
m_t \sim \eta_1^{(0)} = \eta_2^{(0)} = 0.01 \;, \\ 
m_s \sim \eta_3^{(0)} = -0.2 \;, 
\end{array} 
\end{equation} 
where $\eta^{(0)}$'s are bare-couplings as the initial values of the 
differential equations (\ref{dfeq_m1}): $\eta^{(0)} \equiv \eta(L=1)$. 

\begin{figure}[ht]
\begin{picture}(0,270)
\leavevmode\centering\includegraphics{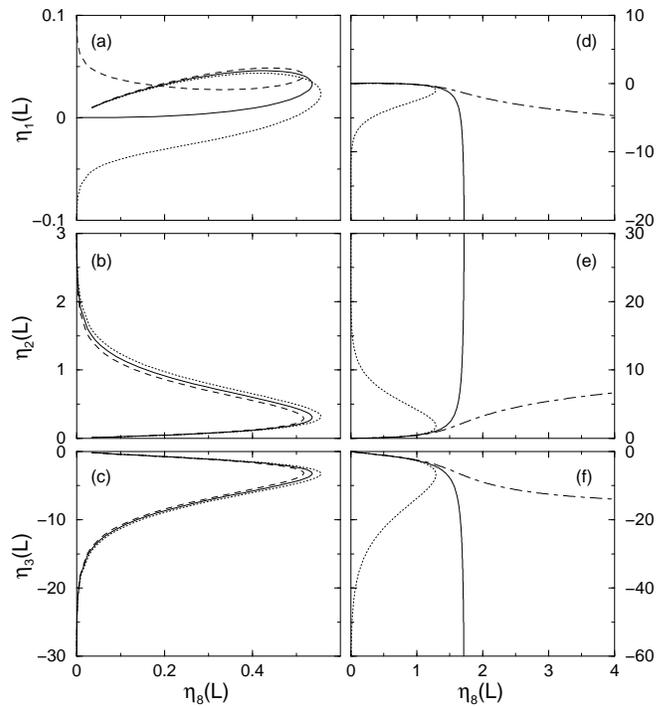}
\end{picture}
\caption{Parametric plot of the RG flows across two phase transitions. The 
figures in the left column describe the flows in the vicinity of the U(1) 
transition, where the solid lines correspond to $\eta_8^{(0)} 
= (\eta_8^{(0)})_{c1}$, the dashed lines to $\eta_8^{(0)}= (\eta_8^{(0)})_{c1}
-0.001$, and the dotted lines to $\eta_8^{(0)}= (\eta_8^{(0)})_{c1} +0.001$. 
The figures in the right column are in the vicinity of the second
(unidentified) transition, where the solid lines correspond to $\eta_8^{(0)} 
= (\eta_8^{(0)})_{c2}$, the dotted lines to $\eta_8^{(0)}= (\eta_8^{(0)})_{c2}
-0.001$, and the dash-dotted lines to $\eta_8^{(0)}= (\eta_8^{(0)})_{c2} +0.001$. 
Notice in panel (f) the dash-dotted line for $\eta_3(L)$ eventually 
approaches $+\infty$ after an upturn that is not shown in the figure.} 
\label{fig_md1_u1foflow} 
\end{figure}

Now the system is subjected to a staggered field. It is found when the 
field is increased to 
\begin{equation} 
h_{c1} \sim (\eta_8^{(0)})_{c1} = 0.0354863 \;, 
\end{equation} 
the system undergoes a second-order phase transition with a U(1) universality 
class (or Gaussian fixed point). The behaviors of the RG flows in the vicinity 
of this transition is shown in Figs.~\ref{fig_md1_u1foflow} (a), (b), and (c).  
Below the U(1) transition the system is in the bulk massive phase B:~H$^{(+)}$ 
characterized by the couplings renormalized to $(\eta_1^{(\infty)}, 
\eta_2^{(\infty)}, \eta_3^{(\infty)}, \eta_8^{(\infty)}) = (+\infty, +\infty, 
-\infty, 0)$ (the dashed lines in the figures). At the transition, the 
infrared property becomes $(\eta_1^{(\infty)}, \eta_2^{(\infty)}, 
\eta_3^{(\infty)}, \eta_8^{(\infty)}) = (0, +\infty, -\infty, 0)$ 
(the solid lines in the figures). Since $\eta_1$ is proportional to the split 
doublet mass, $m_t^d$, it is bound to attribute this transition to a U(1) 
character. Now we are in the bulk phase of B:~I characterized by 
$(-\infty, +\infty, -\infty, 0)$ (the dotted lines in the figures). This is 
an intermediate phase interpolating between the Haldane phase and the 
spontaneously dimered phase. The most distinctive features of this phase are 
that it has only transverse coherent spin excitations and longitudinal string 
order parameter. (For detailed properties, cf. Ref.~\onlinecite{wefn02}.)

If the staggered field is further increased to reach the point 
\begin{equation} 
h_{c2} \sim (\eta_8^{(0)})_{c2} = 0.0560487 \;, 
\end{equation} 
the second transition occurs. The RG flows around this transition is shown in 
Figs.~\ref{fig_md1_u1foflow} (d), (e), and (f). At the transition the 
couplings are renormalized to $(-\infty, +\infty, -\infty, \sqrt{3})$ 
(the solid lines in the figures). After the transition the system enters the 
B:~H$^{(-)}$ phase characterized by $(-\infty, +\infty, +\infty, \infty)$ 
(the dash-dotted lines in the figures). 
Unfortunately, we can tell very little from the present scheme regarding the 
property of this transition. It may be a first-order transition or even a 
crossover. 

In this way, we are able to classify most of the phases and transitions by 
the infrared properties inferred from the RG equations. We summarize the main 
results as follows. 

\paragraph*{a. Bulks}:
\begin{equation} 
\begin{tabular}{l|cccc} 
\hline \hline
& $\eta_1^{(\infty)}$ & $\eta_2^{(\infty)}$ & $\eta_3^{(\infty)}$ 
& $\eta_8^{(\infty)}$ \\ \hline 
B:sD$^{(+)}$ & $+\infty$ & $+\infty$ & $+\infty$ & $0$ \\ 
B:sD$^{(-)}$ & $-\infty$ & $-\infty$ & $-\infty$ & $0$ \\ 
B:H$^{(+)}$ & $+\infty$ & $+\infty$ & $-\infty$ & $0$ \\ 
B:H$^{(-)}$ & $-\infty$ & $+\infty$ & $+\infty$ & $\infty$ \\ 
B:I & $-\infty$ & $+\infty$ & $-\infty$ & $0$ \\ \hline \hline 
\end{tabular}
\label{m1blk} 
\end{equation} 
Comparing with the zero-field case (previous subsection), we find that 
three massive phases sD$^{(\pm)}$ and H$^{(+)}$ are essentially passed 
from their $h=0$ counterparts directly except for the broken symmetry $\eta_1 
\ne \eta_2$. However, B:~H$^{(-)}$ phase is different from S:~H$^{(-)}$ in the 
sense that $\eta_2^{(\infty)} \to +\infty$ instead of $-\infty$ and 
$\eta_8^{(\infty)} \to \infty$ instead of $0$. Obviously, B:~I is a totally 
new phase which has no zero-field counterpart. 

\paragraph*{b. Sheets}: 
\begin{equation} 
\begin{tabular}{l|cccc} 
\hline \hline 
& $\eta_1^{(\infty)}$ & $\eta_2^{(\infty)}$ & $\eta_3^{(\infty)}$ 
& $\eta_8^{(\infty)}$ \\ \hline 
S:U(1) [H$^{(+)}$-I] & $0$ & $+\infty$ & $-\infty$ & $0$ \\ 
S:Z$_2$ [sD$^{(-)}$-I] & $-\infty$ & $0$ & $-\infty$ & $0$ \\ 
S:Z$_2$ [H$^{(+)}$-sD$^{(+)}$] & $+\infty$ & $+\infty$ & $0$ & $0$ \\ 
S:Z$_2$ [sD$^{(-)}$-H$^{(-)}$] & $-\infty$ & $-\infty$ & $-\infty$ & $\sqrt{3}$ \\ 
S: [I-H$^{(-)}$] & $-\infty$ & $+\infty$ & $-\infty$ & $\sqrt{3}$ \\ 
S: [H$^{(+)}$-H$^{(-)}$] & $+\infty$ & $+\infty$ & $-\infty$ & $\sqrt{3}$ \\ 
S: [sD$^{(+)}$-H$^{(-)}$] & $+\infty$ & $+\infty$ & $+\infty$ & $\sqrt{3}$ 
\\ \hline \hline  
\end{tabular}
\end{equation} 
It is perspicuous that the critical surfaces split from the SU(2)$_2$ critical 
line and having U(1) and Z$_2$ criticalities correspond to the infrared 
fixed points with $(\eta_1, \eta_8) \propto (m_t^d, h) \to (0,0)$ and 
$(\eta_2, \eta_8) \propto (m_t^s, h) \to (0,0)$, respectively. The other Z$_2$ 
criticality is trivially same as and connected to the one when $h=0$, 
$(\eta_3, \eta_8) \propto (m_s, h) \to (0,0)$. At least one of the four 
unknown boundaries (dashed lines in Fig.~\ref{fig_pd_md1} and 
\ref{fig_md1_sec}) can be identified as Z$_2$, i.e., S: [sD$^{(-)}$-H$^{(-)}$]. 
The reason is to be shown soon. 

\paragraph*{c. Lines}:
\begin{equation} 
\begin{tabular}{l|cccc} 
\hline \hline 
& $\eta_1^{(\infty)}$ & $\eta_2^{(\infty)}$ & $\eta_3^{(\infty)}$ 
& $\eta_8^{(\infty)}$ \\ \hline 
L:U(1) [H$^{(+)}$-I-H$^{(-)}$] & $+\frac{3}{4}$ & $+\infty$ & $-\infty$ & $\sqrt{3}$ \\ 
L:Z$_2$ [sD$^{(-)}$-I-H$^{(-)}$] & $-\infty$ & $-\frac{3}{4}$ & $-\infty$ & $\sqrt{3}$ \\ 
L:Z$_2$ [H$^{(+)}$-sD$^{(+)}$-H$^{(-)}$] & $+\infty$ & $+\infty$ &$-\frac{3}{4}$ & $\sqrt{3}$ \\ 
\hline \hline 
\end{tabular}
\end{equation} 
These lines are intersections of two surfaces. The special values here are 
associated with the ``fixed point'' (b) of Eqs.~(\ref{dfeq_m1}). 

\begin{figure}[ht]
\begin{picture}(0,170)
\leavevmode\centering\includegraphics{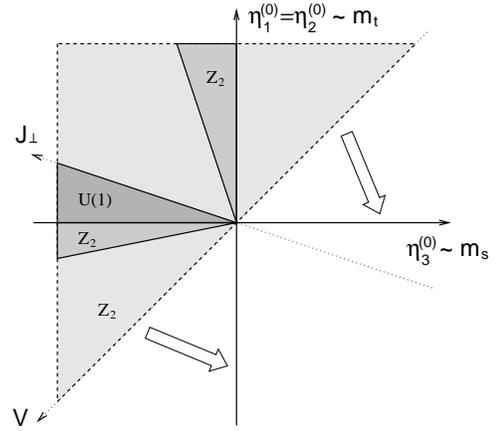}
\end{picture}
\caption{A top-view plot of the phase diagram of the generalized spin ladder 
in a staggered field. The true phase boundaries in the base ($h=0$) should be 
extended as shown by the arrows.} 
\label{fig_md1_bas} 
\end{figure} 
To close our discussion, we show in Fig.~\ref{fig_md1_bas} a top-view of the 
phase diagram. The U(1) zone is confined in a wedged area demarcated by the 
SU(2)$_2$ line $\eta_1^{(0)} = \eta_2^{(0)} =0$ and $\eta_1^{(0)} = 
\eta_2^{(0)} = -\frac{1}{3} \eta_3^{(0)}>0$, 
i.e., $m_t= -\frac{1}{3} m_s>0$. The latter condition amounts to $V=0$ and 
$J_\perp >0$ via relation (\ref{mrjv}), coinciding with the location of a 
standard spin ladder with antiferromagnetic interchain coupling. The adjacent 
Z$_2$ region is delimited by condition $\eta_1^{(0)} = \eta_2^{(0)} = 
\frac{1}{5} \eta_3^{(0)}<0$. The trivial Z$_2$ region is within the Z$_2$ line 
$\eta_3^{(0)} = 0$ and  $\eta_1^{(0)} = \eta_2^{(0)} = 
- 3 \eta_3^{(0)}>0$. The unidentified sheet have $\eta_1^{(0)} = \eta_2^{(0)} 
= \eta_3^{(0)}$ as its boundary in the base. This is a model with $J_\perp=0$, 
i.e., only four-spin interchain coupling. However, one might have noticed that 
there is something unnatural here. Considering the internal relation between 
the B:~H$^{(-)}$ phase in the bulk and the S:~H$^{(-)}$ phase in the base, we 
suggest in the true situation the boundary of this sheet should bend over to 
connect to the critical lines in the base (as shown by the arrows in 
Fig.~\ref{fig_md1_bas}). By combining the ``integrating out'' method and the 
study of the bosonized model, we observe that a Z$_2$ surface, connecting two 
axes, covers the whole third quadrant in Fig.~\ref{fig_md1_bas}.

\subsection{Phase diagram of model (II)} 

Now we turn to Model (II), the generalized spin ladder (\ref{ham_gen}) under 
dimerization (\ref{ham_dlt}). It follows from Eqs.~(\ref{rgm2}) by retaining 
only the relevant couplings:
\begin{eqnarray} 
\dot{\eta}_1 &=& \eta_1 - \frac{1}{4} \eta_8^2 \;, \nonumber \\ 
\dot{\eta}_3 &=& \eta_3 - \frac{1}{4} \eta_8^2 \;, \nonumber \\ 
\dot{\eta}_8 &=& \frac{3}{2} \eta_8 - \frac{1}{2} (3\eta_1 + \eta_3) \eta_8 \;. 
\label{dfeq_m2}
\end{eqnarray} 
This is a model with an unbroken O(3) symmetry. As before, we have an invariance 
$\eta_8 \to -\eta_8$ (i.e., $\Delta \to -\Delta$). In contrast with the previous 
model, now the system has really a fixed point $(\eta_1^*, \eta_3^*, \eta_8^*) 
= (\frac{3}{4}, \frac{3}{4}, \sqrt{3})$ besides the trivial one at 
the origin. This fixed point occurs at $\eta_1 =\eta_3$, or equivalently 
$m_t =m_s$, corresponding to the pure four-spin coupling model ($J_\perp =0$) 
when the system has a higher SU(2)$\times$SU(2)$\approx$O(4) symmetry. This is 
also the condition for the line where the SU(2)$_1$ sheet merges the Z$_2$ 
sheet into the SU(2)$_2$ sheet. The RG flows around this fix point are shown in 
\begin{figure}[ht]
\begin{picture}(0,145)
\leavevmode\centering\includegraphics{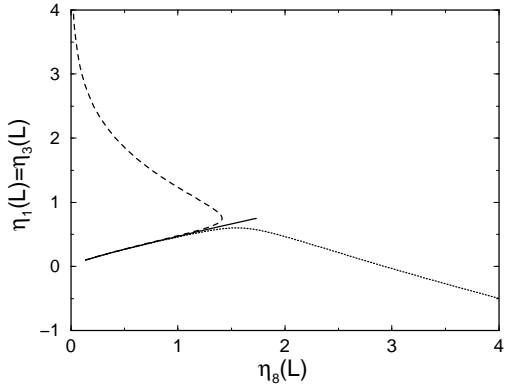}
\end{picture}
\caption{RG flows to the fixed point $(\eta_1^*, \eta_3^*, \eta_8^*) 
= (\frac{3}{4}, \frac{3}{4}, \sqrt{3})$ (solid line) at $\eta_1^{(0)} 
= \eta_3^{(0)}=0.1$ and $(\eta_8^{(0)})_c = 0.137410$, and below 
($\eta_8^{(0)} = (\eta_8^{(0)})_c -0.001$, dashed line) and above 
($\eta_8^{(0)} = (\eta_8^{(0)})_c +0.001$, dotted line) the transition.}
\label{fig_md2_sfl} 
\end{figure} 
Fig.~\ref{fig_md2_sfl}, where we choose the initial values for the base quantities: 
$\eta_1^{(0)} = \eta_3^{(0)}=0.1$. The initial value of the dimerization for 
the fixed point is $(\eta_8^{(0)})_c = 0.137410$. 

\begin{figure}[ht]
\begin{picture}(0,145)
\leavevmode\centering\includegraphics{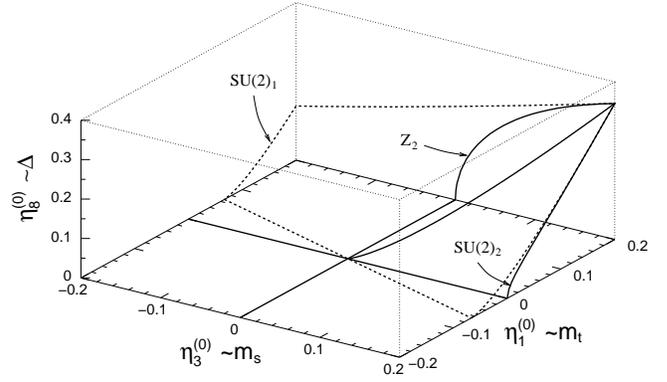}
\end{picture}
\caption{Phase diagram of the generalized spin ladder under $\pi$-phase 
dimerization.}
\label{fig_pd_md2} 
\end{figure}
\begin{figure}[ht]
\begin{picture}(0,260)
\leavevmode\centering\includegraphics{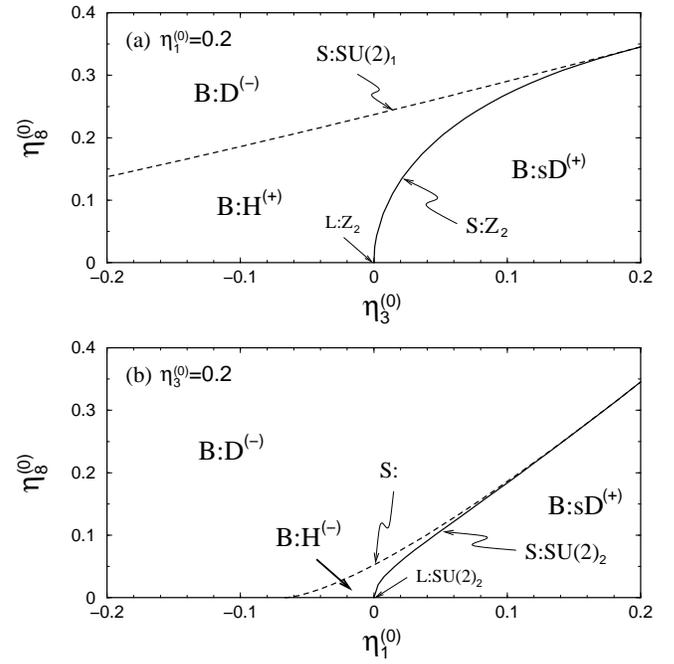}
\end{picture}
\caption{Sectional plot of the phase diagram of the generalized spin 
ladder under dimerization.} 
\label{fig_md2_sec} 
\end{figure}

 
The phase diagram of this model is plotted in Fig.~\ref{fig_pd_md2}, and 
Fig.~\ref{fig_md2_sec} is for the sectional pictures of the same phase 
diagram, where we have labeled various phases. 
As before, the classification of phases and criticalities is associated with 
the infrared properties of the system resulted from the RG equations. 

\paragraph*{a. Bulks}:
\begin{equation} 
\begin{tabular}{l|ccc} 
\hline \hline
& $\eta_1^{(\infty)}$ & $\eta_3^{(\infty)}$ 
& $\eta_8^{(\infty)}$ \\ \hline 
B:H$^{(+)}$ & $+\infty$ & $-\infty$ & $0$ \\ 
B:H$^{(-)}$ & $-\infty$ & $+\infty$ & $0$ \\ 
B:sD$^{(+)}$ & $+\infty$ & $+\infty$ & $0$ \\ 
B:D$^{(-)}$ & $-\infty$ & $-\infty$ & $\infty$ \\ \hline \hline 
\end{tabular}
\end{equation} 
Three massive phases H$^{(\pm)}$ and sD$^{(+)}$ are same as their $\Delta=0$ 
counterparts. Since the symmetry of phase sD$^{(+)}$ is different from the 
explicit dimerization perturbation (\ref{ham_dlt}), it is still spontaneously 
dimerized. However, B:~D$^{(-)}$ phase is different from S:~sD$^{(-)}$, because 
now the system is explicitly dimerized with the same symmetry of the perturbed 
Hamiltonian. In the infrared limit $\eta_8^{(\infty)} \to \infty$ instead of $0$.
It is also noticed from the figures that phase B:~H$^{(-)}$ is restricted to a 
rather small region in the phase space. This may not be the real situation (see 
discussions at the end of this section). 

\paragraph*{b. Sheets}: 
\begin{equation} 
\begin{tabular}{l|ccc} 
\hline \hline 
& $\eta_1^{(\infty)}$ & $\eta_3^{(\infty)}$ & $\eta_8^{(\infty)}$ \\ \hline 
S:SU(2)$_2$ [sD$^{(+)}$-H$^{(-)}$] & $0$ & $+\infty$ & $0$ \\ 
S:Z$_2$ [sD$^{(+)}$-H$^{(+)}$] & $+\infty$ & $0$ & $0$ \\ 
S:SU(2)$_1$ [H$^{(+)}$-D$^{(-)}$] & $+\infty$ & $-\infty$ & $\sqrt{3}$ \\ 
S: [H$^{(-)}$-D$^{(-)}$] & $-\infty$ & $+\infty$ & $\sqrt{3}$ \\ \hline \hline  
\end{tabular}
\end{equation} 
Like the case before, the identification of the SU(2)$_2$ and Z$_2$ 
criticalities is directly due to the fact that in addition to the dimerization 
($\eta_8^{(\infty)}$), the triplet ($\eta_1^{(\infty)}$) and singlet 
($\eta_3^{(\infty)}$) masses are renormalized to vanish, respectively. 
However, the reference of the interface between phases H$^{(+)}$ and D$^{(-)}$ 
to an SU(2)$_1$ criticality is a bit subtle. This conclusion is drawn from 
other studies of the standard ladder under 
dimerization.\cite{tnhs95,ms9698,cg99,wn00} In Ref.~\onlinecite{tnhs95}, a 
standard spin-$\frac{1}{2}$ ladder with ferromagnetic interchain coupling 
under $0$-phase dimerization (the dual case of our present model) was 
semiclassically mapped onto an O(3) nonlinear sigma model with some 
topological angle. The SU(2)$_1$ criticality is realized when this angle 
equals $\pi$. On the other hand, an effective Hamiltonian was explicitly 
derived in Ref.~\onlinecite{wn00} which describes the same transition in terms 
of a single spin-$\frac{1}{2}$ chain with bond alternation, and the 
transmutation of all physical quantities from the ultraviolet fixed point to 
the infrared fixed point was established. We have only one unidentified 
surface, S: [H$^{(-)}$-D$^{(-)}$], in this model. 

\paragraph*{c. Line}:
\begin{equation} 
\begin{tabular}{l|ccc} 
\hline \hline 
& $\eta_1^{(\infty)}$ & $\eta_3^{(\infty)}$ & $\eta_8^{(\infty)}$ \\ \hline 
L:SU(2)$_2$ [sD$^{(+)}$-H$^{(+)}$-D$^{(-)}$-H$^{(-)}$] & $\frac{3}{4}$ 
& $\frac{3}{4}$ & $\sqrt{3}$ \\ \hline \hline 
\end{tabular}
\end{equation} 
The only critical line in this model represents the fixed point of 
Eqs.~(\ref{dfeq_m2}), the RG flow for the point on which is shown in 
Fig.~\ref{fig_md2_sfl}. This is a very special line with a higher O(4) 
symmetry: it is not only the ridge line where all sheets meet, but also 
the common boundary shared by all four phases (blocks) of the model. 

\begin{figure}[ht]
\begin{picture}(0,170)
\leavevmode\centering\includegraphics{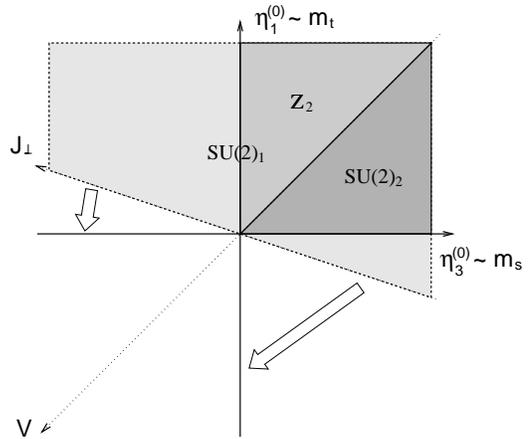}
\end{picture}
\caption{Top-view plot of the phase diagram of the generalized spin ladder 
under dimerization. The true SU(2)$_1$ sheet should connect to the SU(2)$_2$ 
line in the base, and the unidentified sheet to the Z$_2$ line in the base 
(as shown by the arrows).} 
\label{fig_md2_bas} 
\end{figure} 
Since the obtained phase diagram is grounded on the relevant part of the 
one-loop RG equations, the position of the phase boundaries may not be so 
correct. For instance, considering the intimacy between the bulk B:~D$^{(-)}$ 
phase and the S:~sD$^{(-)}$ phase in the base, most plausibly the SU(2)$_1$ 
sheet is connected to the SU(2)$_2$ line of $\Delta =0$, and the boundary for 
the unidentified sheet is pushed to connect to the Z$_2$ line in the base (see 
Fig.~\ref{fig_md2_bas}). So the space for the bulk B:~H$^{(-)}$ phase is 
considerably enlarged. This is supported by the fact that, as we have 
mentioned, the ($J_\perp >0$) standard ladder ($V=0$) undergoes the SU(2)$_1$ 
transition under the ($\pi$-phase) dimerization. So the SU(2)$_1$ sheet cannot 
terminate at the location which is a result from the approximate RG 
Eqs.~(\ref{dfeq_m2}) as shown in Fig.~\ref{fig_pd_md2}.

\section{Summary and Discussions} \label{sec_sum}

Based on the one-loop RG equations and the assumption that the relevant 
perturbations govern the relative structure of various  phases in the 
parameter space, we have speculated from the infrared properties of these 
equations the phase diagrams of two related spin ladder models in the 
weak-coupling regime. 

In model (I), the external staggered field breaks the SU(2) symmetry down to 
U(1) and Z$_2$, so that the criticalities at most have the same symmetries 
and no room for SU(2) to live. Indeed, we find both U(1) and Z$_2$ 
criticalities come forth with the introduction of the staggered field (see 
Fig.~\ref{fig_pd_md1} and \ref{fig_md1_bas}), which actually evolve from the 
underlying SU(2)$_2$ criticality in absence of the field. The present result 
confirms the picture we obtained through a somewhat different approach in 
Ref.~\onlinecite{wefn02}. In that study, we literally integrated out the fast 
degrees of freedom associated with the singlet modes close to the SU(2)$_2$ 
line and derived an effective action which describes the triplet sector with 
anisotropy induced by the staggered field. The vanishing of the split 
effective triplet mass manifests the criticality accordingly. 
The new observation is that the intermediate phase (B:~I), which interpolates 
between the Haldane phase (B:~H$^{(+)}$) and the dimer phase (B:~sD$^{(-)}$), 
is restricted to a very limited region, and there should exist another 
(unidentified) transition which separates B:~I phase from the large-$h$ 
(Haldane) phase (B:~H$^{(-)}$) which occupies most of the space in the phase 
diagram. From the infrared property (\ref{m1blk}), we see that B:~I phase is 
quite unique entirely due to the staggered field, and there is no resemblance 
preexisting when $h=0$. 

Comparing with model (I), model (II) is relatively simple in the sense that 
the SU(2) symmetry is unbroken and the O(3) symmetry in the triplet sector 
remains, and the number of RG equations is thus reduced. All four bulk phases 
can basically be traced back to their counterparts when $\Delta =0$. None the 
less, as we have seen, the phase diagram of this model is by no means 
uneventful (see Fig.~\ref{fig_pd_md2} and \ref{fig_md2_bas}). The phenomenon 
extremely intriguing is the merging of SU(2)$_1$ and Z$_2$ critical sheets 
into SU(2)$_2$ sheet. The information encoded in the RG equations reveals that 
this occurs at some critical dimerization strength $\Delta_c(V)$ when the 
system has a higher O(4) symmetry or SU(2)$\times$SU(2) in the original 
lattice model, i.e., $J_\perp =0$. The merging line with SU(2)$_2$ criticality 
is shared by all four massive phases of the model. 

The benevolent aspect of our present theory is that we are able to assort 
different phases according to their infrared properties from the RG flows, 
and the identification of the critical sheets that emerge from the underlying 
critical lines in the base should be reliable. These critical sheets, 
including two Z$_2$ and U(1) in model (I) and SU(2)$_2$ and Z$_2$ in model 
(II), can also be verified to exist\cite{nwup} by means of ``integrating out'' 
procedure employed in Ref.~\onlinecite{wefn02}, although the terminating 
positions of these sheets are unknown by this approach. However, as one may 
have perceived, the insolvency of the present theory is that not all of the 
transitions can be attributed. Even worse is the phase boundaries resolved 
from the one-loop equations may not be accurate. (See our suggestions for 
the modification of the phase diagrams in Fig.~\ref{fig_md1_bas} and 
\ref{fig_md2_bas}.) Therefore, although the topology of the phase diagrams 
we believe has been overall and qualitatively captured, the unwavering 
pin-pointing of the phase diagrams has to resort to other, say numerical, 
methods. For example, it is predicted here for model (II) of pure four-spin 
interaction ($V$), when increasing the dimerization, there is only one 
second-order (SU(2)$_2$) transition which connects directly two dimerized 
phases with mutually mismatched orders. If adding a bit positive 
$J_\perp$ coupling, one would expect two consecutive transitions: Z$_2$ and 
SU(2)$_1$ in a row.

\acknowledgments

It is a pleasure to acknowledge the helpful discussions with P.\ Fulde, 
K.\ Penc, R.\ Narayanan, T.\ Vekua, and M.\ Nakamura. Special thanks are due 
to A.\ Nersesyan for calling the problem to my attention. This work was 
supported under the visitors program of MPI-PKS. 



\begin{thebibliography}{99}
\bibitem{note1} Rigorously speaking, the interchain coupling $J_\perp$ shifts 
the triplet and singlet velocities differently. However, we neglect this 
difference under the condition (\ref{wccnd}).
\bibitem{note2} The definitions and conventions we take may be found, e.g., 
in Appendix A of Ref.~\onlinecite{wefn02}.
\end{thebibliography}
\end{document}